\documentclass[twocolumn,showpacs]{revtex4}
\usepackage{epsf,graphicx}
\usepackage{subfigure}
\usepackage{amsmath,amssymb,amsfonts}
\begin{document}
   \newcommand{\cent}[1] {\begin{center}#1\end{center}}
   \newcommand{\doublint}{\int\rule{-3.5mm}{0mm}\int}
   \newcommand{\mra}  {\to}
   \newcommand{\vecbm}[1]{\mbox{$\boldmath#1$}}
   \newcommand{\loo}{\,\raisebox{-.5ex}{$\stackrel{<}{\scriptstyle\sim}$}\,}
   \newcommand{\goo}{\,\raisebox{-.5ex}{$\stackrel{>}{\scriptstyle\sim}$}\,}
   \newcommand{\lra}  {$\leftrightarrow$}
   \newcommand{\vecb}[1]{\mbox{\bf#1}}
\paperwidth =15cm 
\title{Comment on ``Negative heat and first order phase transitions in
  nuclei''\\ by Moretto et al.}  \author{D.H.E. Gross} \address{
  Hahn-Meitner-Institut Berlin, Bereich Theoretische
  Physik,Glienickerstr.100\\ 14109 Berlin, Germany and Freie
  Universit{\"a}t Berlin, Fachbereich Physik; \today}
\begin{abstract}
  The recent paper nucl-th/0208024 by Moretto et al. is commented:
  Their picture of nuclear phase transition in terms of macroscopic
  control parameters, temperature and pressure, is irrelevant. Their
  criticism of order-disorder phase-transitions on a periodic lattice
  uses the wrong scenario. This transition has nothing to do with the
  liquid-gas transition of a single spherical droplet.
\end{abstract}
\maketitle
In their recent paper (nucl-th/0208024) Moretto et.al. come up with
the statement that negative heat capacities in calculations on a
lattice (Ising or Potts-model) with periodic boundaries are
artificial. This continues an old and long lasting dispute I have with
Moretto c.f.~\cite{gross146,gross179} about his macroscopic picture
of a microscopic process.

Here some additional remarks on our lattice simulations of the caloric
curve in the $q=10$ Potts model~\cite{gross150}:

\begin{itemize}
\item In these lattice calculations the transition from ordered to
  disordered spin-distributions at constant volume is discussed, not
  the evaporation of a single drop at constant pressure.
\item Therefore, in a completely filled lattice at phase-separation
  are one to several regions with disordered spins (``bubbles'') inside
  regions of ordered (predominantly parallel) spins. At larger
  excitation we have the opposite situation of phase-separation:
  several ``droplets'' of ordered spins in a disordered background.
\item The interface surfaces are not at all neccessarily the surface
  of a single spherical droplet. Only the microcanonical ensemble
  allows to follow the details of this methamorphosis from the ordered
  phase to the disordered phase. Intensive variables as pressure and
  temperature, used by Moretto are useless as control-parameters at
  phase separation in small systems and control the thermodynamics
  only in the thermodynamic limit {\em and only in pure phases far
    away from any transition.}  The constructions in fig. 3 of Moretto
  et.al.  paper are irrelevant.
\item At finite excitation {\em the surfaces have of course an
    entropy} connected to fluctuations in their shape and in their
  number (i.e.  creation of several bubbles/droplets). This surface
  entropy leads to the diminishing of the surface tension which
  finally becomes zero at the critical point. Moretto's figure fails
  to catch this important physics.
\item And again, even if Moretto claims the contrary: {\em Nuclear
    multifragmentation} is not the story of a single nuclear drop
  embedded in its vapor at some given temperature and pressure.  In
  nuclear fragmentation {\em experiments} there is a constant energy
  but neither a heat bath (temperature) nor a vapor pressure. I. e.
  the realistic scenario is not as trivial and macroscopic as Moretto
  et al. claim.
\item In our calculations of {\em nuclear}
  fragmentation~\cite{gross176,gross95} we did of course {\em not} use
  periodic boundary conditions but a fixed freeze-out volume of about
  $\sim 4-6$ times normal nuclear volume. The physical scenario I had
  in mind is fragments separating from their original source
  experience a {\em strong and short-ranged frictional coupling} as is well
  known from deep-inelastic collisions. This may likely lead to equilibration
of the already fragmented configuration in a quite narrow freeze-out volume
as it was assumed in my $MMMC$ model~\cite{gross95}.
\end{itemize}


\end{document}